\documentclass[jap, reprint]{revtex4-1}

\usepackage{amsmath, amssymb}
\usepackage{graphicx}
\usepackage{dcolumn}
\usepackage{bm}

\usepackage[utf8]{inputenc}
\usepackage[T1]{fontenc}
\usepackage{mathptmx}
\usepackage{etoolbox}

\makeatletter
\def\@email#1#2{%
 \endgroup
 \patchcmd{\titleblock@produce}
  {\frontmatter@RRAPformat}
  {\frontmatter@RRAPformat{\produce@RRAP{*#1\href{mailto:#2}{#2}}}\frontmatter@RRAPformat}
  {}{}
}
\makeatother

\begin{document}

\title{High power vacuum ultraviolet source based on gasdynamic ECR plasma}

\author{R. L. Lapin}
    \email{lapin@ipfran.ru}
\author{V. A. Skalyga}
\author{S. V. Golubev}
\author{I. V. Izotov}
\author{S. V. Razin}
    \affiliation{Institute of Applied Physics of Russian Academy of Sciences, 46 Ul'yanov Street, Nizhny Novgorod, 603950, Russia}

\author{O. Tarvainen}
    \affiliation{STFC ISIS Pulsed Spallation Neutron and Muon Facility, Rutherford Appleton Laboratory, Harwell, OX11 0QX, UK}

\date{\today}

\begin{abstract}
We report experimental results of vacuum ultraviolet (VUV) emission from the plasma of an electron cyclotron resonance (ECR) discharge in hydrogen, sustained by powerful millimeter-wavelength radiation of a gyrotron. Distinctive features of the considered discharge are the high plasma density ($10^{13}$ cm$^{-3}$ order of magnitude) and, at the same time, the high electron average energy (10 --- 300 eV), which makes it possible to significantly increase the efficiency of VUV re-emission of the energy deposited into the plasma by the microwave radiation. Experiments were performed with the plasma confined in a simple mirror trap and heated by pulsed gyrotron radiation 37.5 GHz / 100 kW under the ECR condition. The measured volumetric VUV emission power of Lyman-alpha line (122 ± 10 nm) overlapping with the Werner band, Lyman band (160 ± 10 nm), and molecular continuum (180 ± 20 nm) reached 45, 25, and 55 W/cm$^3$, respectively. The total absolute radiation power in these three ranges integrated over the plasma volume is estimated to be 22 kW i.e. 22\% of the incident microwave power, which matches theoretical predictions. Further optimization of the conditions of the ECR hydrogen discharge sustained by powerful gyrotron radiation provides an opportunity for the development of effective technological VUV sources of a kilowatt power level.
\end{abstract}


\maketitle

\section{Introduction}

Ultraviolet radiation is used in numerous industrial applications. Some of those require further development of existing UV sources. One of the most trending technologies is solid-state electronics based on synthetic diamonds. Diamond is a unique material with the highest thermal conductivity amongst solids, and high electrical strength and drift velocity ($>3\times10^3$ cm$^2$V$^{-1}$s$^{-1}$) \cite{Schneider}. Diamond-based electronic devices (e.g.., the control of diamond switches\cite{Calvani}) are based on producing a high density of free charge carriers in the conduction band due to irradiation by photons with their energy exceeding the 5.5 eV ($\lambda\approx 225$ nm) diamond bandgap \cite{Girolami}. Therefore, high-power vacuum ultraviolet ($\lambda\leq 200$ nm) sources are in high demand for the operation of diamond-based electronic devices designed for power electronics applications. Various UV and VUV sources are used in this field, e.g. gas-discharge Hg (Ar) lamps \cite{Geis} and titanium-sapphire lasers \cite{Yatsenko}, providing emission power in the range from $10^{-6}$ W to 100 W. Yet, further improvement of pulsed and continuous VUV sources up to kilowatt power level would be welcomed by the industry \cite{Park}. Thus, the development of new effective sources of UV-VUV light is an important and challenging research task.

Hydrogen plasmas could be used as a source of the VUV light. They emit VUV-light through electron impact excitation of atoms and molecules and their de-excitation to lower states accompanied by VUV photon emission \cite{Ajello}. Intense VUV-emission can be produced by a range of discharges; inductively or capacitively coupled plasmas, electron cyclotron resonance (ECR) or helicon discharges, etc. According to theoretical estimations \cite{Komppula_th}, the efficiency of the VUV source based on hydrogen discharge can reach 70\% if the ionization fraction of the neutral gas is high. Even at a 1\% ionization fraction, more than 10\% of the incident power can be expected to convert into VUV-radiation \cite{Komppula_th}. Batishchev et al. \cite{Batishchev} studied a helicon discharge by simulations and found that approximately 25\% of the discharge power dissipates via photon emission. While theoretical studies and simulations cover a wide range of plasma parameters and discharge types, experimental investigations of VUV re-emission efficiency of the energy accumulated by the plasma because of its heating by external sources are rarely found in the literature. Komppula et al. \cite{Komppula_exp} measured the volumetric VUV emission power of a 2.45 GHz microwave discharge to be less than $10^{-1}$ W/cm$^3$ with an energy input of several W/cm$^3$. The difference to theoretical estimates in that particular case is attributed to poor microwave plasma coupling \cite{Komppula_exp}.

A promising way towards meeting the above-mentioned estimates is to use an ECR discharge with powerful millimeter-wave heating enabling to increase the plasma density and temperature as well as to miniaturize the plasma characteristic dimension. Even in the case of ECR discharge, there is a wide range of opportunities for VUV emission efficiency tuning. Plasma parameters in such a discharge could vary significantly depending on gas pressure, heating power, and system dimensions.

The most common application of high-frequency (18 --- 45 GHz) ECR discharge is ion sources of multiply charged ions \cite{Geller, Sun}. For that purpose a hot-electron rarefied plasma is produced under low-pressure and high heating power in a minimum-B magnetic trap. In this case, the electron temperature is too high (1 --- 10 keV) and plasma and neutral gas densities ($10^{11}$ --- $10^{12}$ cm$^{-3}$ and $10^{-6}$ Torr, respectively) are too low for the production of a high number of excited hydrogen particles. The situation is dramatically improved in the case of so-called quasi-gasdynamic ECR ion sources. These devices utilize powerful (up to 100 kW) millimeter wave (37 --- 75 GHz) gyrotron radiation for plasma production and reach electron densities up to $10^{14}$ cm$^{-3}$ with 10$^{-3}$ Torr gas pressure. The studies of the plasma properties of gasdynamic ECR sources have been carried out at SMIS 37 facility during the last 2 decades \cite{Golubev04, Skalyga2014, Skalyga2016}. The utilization of 37.5 GHz radiation with pulsed power up to 100 kW for ECR heating of the plasma with a volume of 100 --- 200 cm$^{-3}$ makes SMIS 37 one of the most powerful microwave discharge experiments in terms of volumetric energy input. The unique combination of the system characteristics makes it possible to provide significantly higher electron temperature (10 --- 300 eV) than in 2.45 GHz ECR devices or in inductively coupled plasmas (5 --- 10 eV) even in the case of high electron collision rate, leading to reduced plasma lifetime down to the level determined by the gasdynamic plasma losses from the magnetic trap. The plasma confinement with gasdynamic outflow accompanied by hot electrons with mean free paths comparable or longer to system dimensions is called the quasi-gasdynamic regime.

For a demonstration of the promising features of such plasma for VUV-emission we can make simple estimations: Let us assume electron temperature of 50 eV, electron density of $2\times10^{13}$ cm$^{-3}$, neutral gas pressure of $10^{-3}$ Torr, magnetic trap length, and its mirror ratio of 25 cm and 5, respectively. These are typical values for SMIS 37 experiments \cite{Skalyga2017}. In this case, the plasma lifetime is determined by the particle flow from the trap with ion sound velocity and could be estimated as $\tau_{pl} = LR/V_{is}\approx 1\times10^{-5}$ s, which corresponds to electron loss frequency of about $\nu_{loss} = 10^5$ s$^{-1}$. The excitation frequency of neutral particles to VUV-emitting states $\nu_{exc} = kn_e$ would be $5\times10^5$ s$^{-1}$ and $2.5\times10^5$ s$^{-1}$ for atoms and molecules correspondingly (here k is the rate coefficient calculated from the cross-sections in Ref.~\onlinecite{Janev}). Thus, the excitation frequency is at least a few times higher than the electron loss rate, which implies that in the discharge where the electron energy (50 eV) and the excitation potential (for example, for atoms is 10.2 eV) are on the same order, the VUV emission must be a significant energy loss channel.

The present paper describes experimental studies of VUV emission from the hydrogen plasma at SMIS 37 experimental facility.

\section{Description of the experimental facility and diagnostic methods}

\begin{figure*}
\includegraphics[scale=0.4]{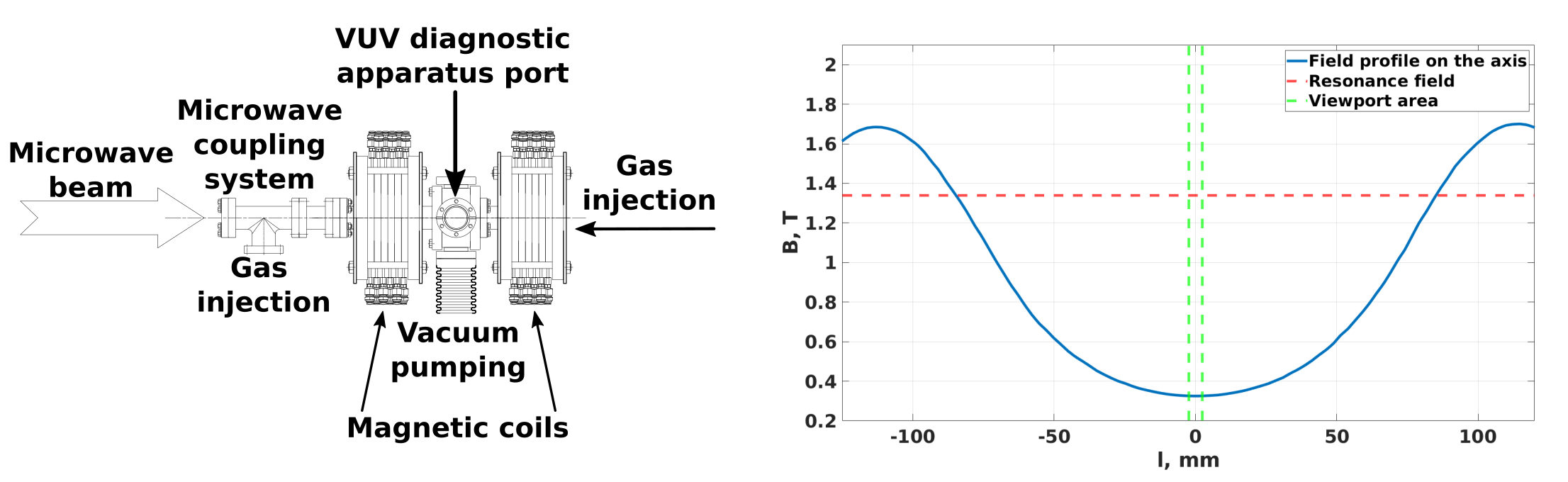}
\caption{\label{fig:scheme} The experimental scheme and magnetic trap axial field distribution together with ECR field mark and photodiode line-of-sight plasma cross-section.}
\end{figure*}

\begin{figure}
\includegraphics[scale=0.25]{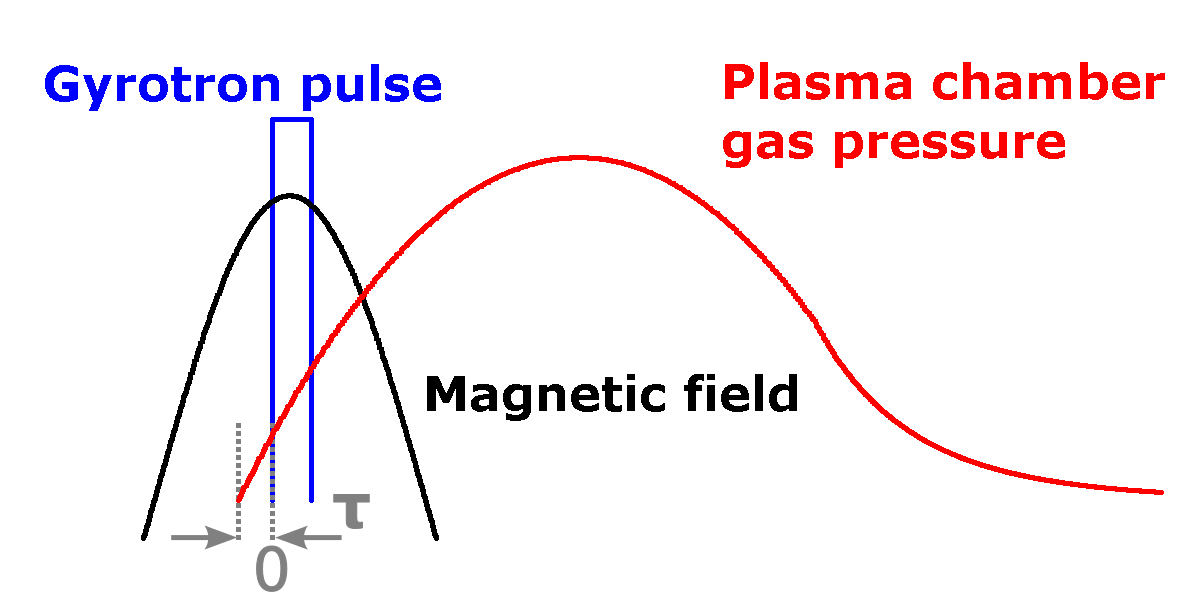}
\caption{\label{fig:temporal} Time structure of the SMIS 37 pulsed operation. $\tau$ is the temporal delay between the leading edges of the gas inlet and microwave pulses.}
\end{figure}

All experiments were performed at the SMIS 37 facility, which is a pulsed ECR plasma and ion source with quasi-gasdynamic plasma confinement, schematically depicted in Fig.~\ref{fig:scheme}. The plasma was produced by gyrotron radiation with 37.5 GHz frequency and the power up to 100 kW was transmitted into the plasma chamber by a quasi-optical matching system. The magnetic trap satisfying the ECR condition was formed by a pair of pulsed water-cooled solenoids. In the experiments described here the magnetic field strength at the trap mirror points was 1.7 T (resonance field for 37.5 GHz is 1.34 T), the distance between the magnetic mirrors was 25 cm and the trap mirror ratio ($B_{mirror}$/$B_{min}$) was 5. The magnetic field distribution is also shown in Fig.~\ref{fig:scheme}. The background pressure was at the level of $10^{-7}$ Torr between the microwave pulses, while the pulsed gas injection was used to reach $10^{-4}$ —-- $10^{-3}$ Torr pressure in the plasma chamber at the moment of the discharge ignition. The pulse repetition rate for the whole experimental facility was 0.1 Hz, thus the gas conditions for consecutive pulses were independent of each other due to efficient plasma chamber vacuum pumping. As far as all main systems of SMIS 37 facility operate in the pulsed mode their time-resolved structure is shown in Fig.~\ref{fig:temporal}. The microwave pulse with a duration of 1 ms was tuned to start in the middle of the magnetic field pulse which had a duration of about 7 ms, providing plasma ignition at the moment of the peak magnetic field of the trap. The gas injection into the plasma chamber was tuned by two parameters: a temporal delay, $\tau$, between the gas valve opening and the microwave pulse leading edge; and the gas feed line pressure, P, upstream of the dosing valve. Thus, the variation of $\tau$ and P allowed to tune the neutral gas pressure at the moment of the discharge ignition and gas mass-flow rate during the microwave pulse. In terms of plasma parameters, $\tau$ and P together with microwave power determine plasma density, neutral gas pressure, ionization degree, and electron temperature.

The absolute power of plasma VUV emission was measured with a calibrated photodiode IRD SXUV5 (Opto Diode Corp.) coupled with a set of retractable optical bandpass filters. Three optical filters were used, corresponding to the three investigated wavelength VUV ranges: Lyman-alpha line of the atomic spectrum ($2P\xrightarrow{}1S$, bandwidth: $122 \pm 10$ nm) overlapping with the Werner band of molecular emission (singlet transition $C^1\Sigma^{+}_u\xrightarrow{}X^1\Sigma^{+}_g$), and two ranges of the molecular spectrum: Lyman band (singlet transition $B^1\Sigma^{+}_u\xrightarrow{}X^1\Sigma^{+}_g$, bandwidth: $160 \pm 10$ nm) and molecular continuum (triplet transition $a^3\Sigma^{+}_g\xrightarrow{}b^3\Sigma^{+}_u$, bandwidth: $180 \pm 20$ nm). The photodiode spectral sensitivity, which was calibrated by the manufacturer, is almost constant in the considered range of 100 –-- 200 nm. The precise sensitivity values taking into account the filters transparencies for all studied wavelength ranges are presented in Table~\ref{tab:efficiency}. Such an optical diagnostics system was assembled in a separate vacuum unit, presented in Fig. 3. A movable filter holder allowed us to change the filter between the diode and the plasma volume. The photodiode sensitive area was 2.5 mm$^2$. The system was equipped with an additional collimator with a 2 mm aperture to avoid any influence of wall reflections and to provide measurements from a known plasma volume. This diagnostic unit was mounted to a radial port viewing the center of the plasma chamber marked in Fig.~\ref{fig:scheme}. The photodiode line-of-sight is schematically shown together with magnetic field distribution in Fig.~\ref{fig:scheme}.

\begin{table}
\caption{\label{tab:efficiency} The efficiency of the IRD SXUV5 (Opto Diode Corp.) photodiode coupled with corresponding bandpass filter for considered wavelength ranges.}
\begin{ruledtabular}
\begin{tabular}{lccc}
Range & Lyman-alpha & Lyman band & Mol. continuum \\
\hline
VUV diagnostic \\ system sensitivity, \\ A/W$\times$cm$^2$ & $2.5\times10^{-5}$ & $4.25\times10^{-5}$ &
$1\times10^{-4}$\\
\end{tabular}
\end{ruledtabular}
\end{table}

\begin{figure}
\includegraphics[scale=0.8]{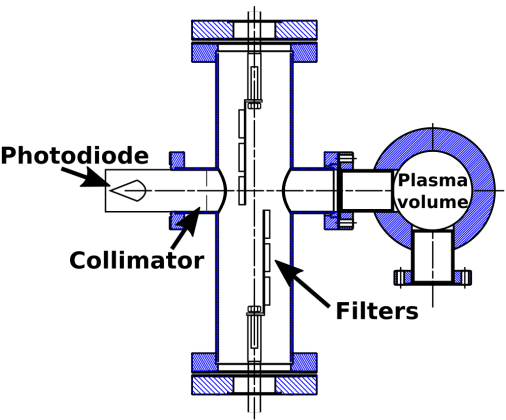}
\caption{\label{fig:VUVsetup} The diagnostic setup scheme for VUV power measurements.}
\end{figure}

The distance from the photodiode to the trap axis was 23.9 cm. The emission volume of the plasma, from which the photons can reach the photodiode, was approximately 0.6 cm$^3$. The total plasma volume can be estimated as the volume limited by the magnetic field lines touching the inner metallic limiters (plasma chamber wall parts) and has a value of about 134 cm$^3$. The ratio between the total plasma volume and the one observed by the diode was used as a factor to calculate the total power of VUV emission produced by the plasma assuming isotropic emission.

Taking into account typical SMIS 37 plasma parameters it can be straightforwardly deduced that the mean free paths of all heavy particles are much smaller than the plasma chamber dimensions, while the electrons have the mean free path comparable to the magnetic trap length but at the same time experience many inelastic and elastic collisions during their lifetime (collisional confinement). The spontaneous lifetime of atomic and molecular excited states emitting in the VUV-range is 1 --- 10 ns (order of magnitude), which means that the VUV photons are emitted right at the spatial location of the electron impact. Because of the small mean free path of the Lyman-alpha photon (can be estimated as a few hundredths of a centimeter according to \cite{Dijkstra} assuming hydrogen pressure and temperature as 1 mTorr and 300 K correspondingly) the opacity of background gas cannot be totally neglected. However, due to subsequent VUV re-emission and the lack of accurate data of atomic density, hydrogen Lyman-alpha transparency is assumed hereinafter. As such the measured VUV-intensity is proportional to the electron impact reaction rate within the line-of-sight volume and there are no metastable states that could transport the chemical potential (otherwise emitting in VUV) away from the discharge volume. The uniform electron distribution along with the trap together with the assumed uniform gas density implies that the VUV-emission can be considered uniform and isotropic, which enables us to calculate the total emission power from the measured volumetric power.

\begin{figure*}
\includegraphics[scale=0.18]{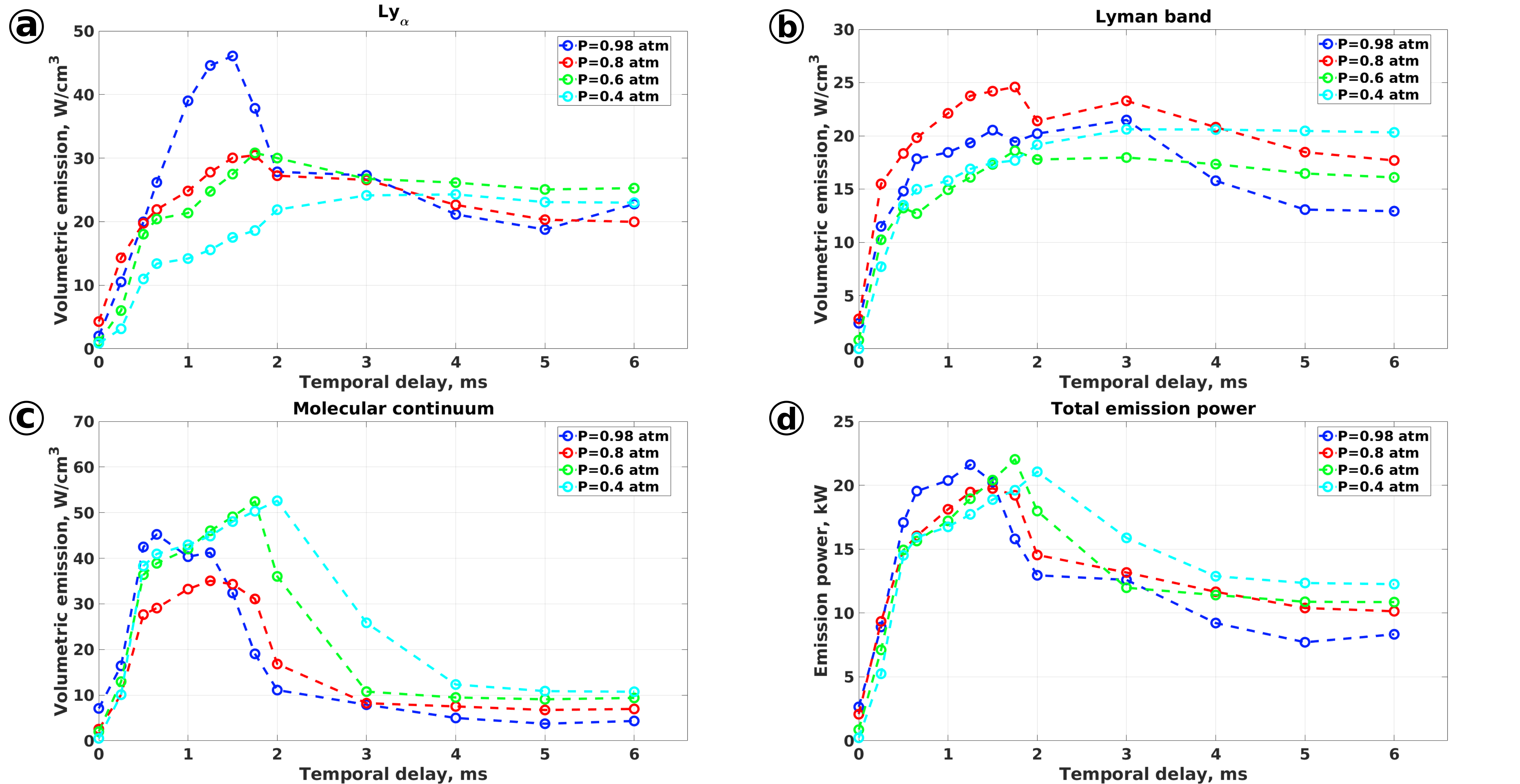}
\caption{\label{fig:results} The dependence of the volumetric VUV power of the a) Lyman-alpha line emission, b) Lyman band emission, c) molecular continuum emission; d) the total absolute VUV emission power on the temporal delay between the gas pulse and microwave pulse.}
\end{figure*}

\section{VUV measurement results}

The dependence of the volumetric VUV emission power (in the three above-mentioned wavelength ranges) on the temporal delay between the gas valve opening and the microwave pulses was measured for a set of gas feed line pressure values. The total absolute VUV emission power (summed over all ranges) was also estimated. The volumetric power of the Lyman-alpha line (affected by the Werner-band emission), the Lyman band, and the molecular continuum are presented in Fig. 4a-c. The data of Fig. 4b-c shows the emission power in the whole Lyman band and molecular continuum. The given values take into account the part of the molecular spectrum outside the bandpass filter transmission range. The estimate is based on synthetic spectra at 0-9000 K vibrational temperature \cite{Komppula_exp_1}.

The maximum measured values of the volumetric VUV power were estimated as 45, 25, and 55 W/cm$^3$ (with a total input power of 100 kW) for the Lyman-alpha line, the Lyman band, and the molecular continuum, respectively. In our estimations, we have not taken into account the influence of the Werner band emission on the Lyman-alpha line because the difference in the corresponding cross-sections is about 10 times for the assumed electron temperature. It is worth noting that contrary to many other low-temperature plasma devices \cite{Komppula_exp, Komppula_exp_1} where Lyman-alpha radiation was found to have the highest power we measured the molecular continuum emission to dominate. The molecular continuum emission corresponds to electronic transitions from the $a^3\Sigma^{+}_g$ triplet state to the repulsive $b^3\Sigma^{+}_u$ state resulting in auto-dissociation of the molecule into ground-state hydrogen atoms. It has been found previously that up to 94\% of the positive ions extracted from the SMIS 37 device are protons (as opposed to molecular ions) \cite{Skalyga2012}, which is commensurate with the emission power of the molecular continuum being very high. The Lyman-alpha radiation reaching the diode detector could be attenuated by absorption of the photons in the hydrogen gas containing protons. Quantifying the effect of the plasma opacity \cite{Behringer} would require estimating the hydrogen atom density on the line-of-sight between the discharge and the detector.

The observed dependencies demonstrate the existence of an optimal range for the temporal delay between the gas valve opening and the microwave pulse. The optimal delay is shorter for higher gas feed line pressures. That obviously means that there is an optimal gas pressure inside the plasma chamber at the moment of gas breakdown which is realized earlier with more intense gas injection. Such an optimum could be explained with a high sensitivity of the electron heating efficiency to gas conditions in the ECR discharge with plasma density around the cut-off value. Just a slight overgrowth above the cut-off usually leads to a strong reflection of heating microwaves from the plasma and a dramatic drop of the electron temperature down to regular 5 --- 10 eV, typical for a variety of usual laboratory discharges. 

It should be noted that the gas feed line pressure was limited to 1 atm, so the highest volumetric VUV power for the Lyman-alpha line was obtained at the maximum considered gas flow rate. That means that probably even better optimization of the discharge conditions and higher VUV emission power could be realized in the future.

Fig. 4d shows the estimated total VUV power. The maximum absolute power of VUV plasma emission was approximately 22 kW. At higher gas feed line pressure the maximum plasma luminosity was reached at a shorter temporal delay between gas valve and microwave pulses. That fact correlates with the data of Fig. 4a-c, demonstrating the importance of gas pressure optimum in the plasma chamber at the moment of the discharge ignition. 

\section{Conclusion}

An experimental study of the ECR discharge plasma sustained by the 37.5 GHz/100 kW radiation as a source of VUV radiation was carried out. It was shown that the use of quasi-gasdynamic hydrogen plasma is a very promising approach as a powerful VUV source. The maximum VUV emission power of 22 kW was measured under the conditions of high neutral hydrogen pressure (of the order of several mTorr in the discharge chamber). Investigations with higher gas injection intensity were not possible due to safety regulation of the maximum pressure for the actual configuration of the gas feed line. Further experiments covering even higher gas pressure values are of great interest for better system optimization.

The measured power accounts for 22\% of the incident power, which is in good agreement with theoretical estimations \cite{Komppula_th}. It is worth noting that as far as measurements were performed with the set bandpass filter the total plasma emission power in the whole VUV range must be somewhat higher. A notable example is the Werner band overlapping with the Lyman-alpha radiation with the electron impact excitation cross-section (from the ground state) to Werner-band emitting states being comparable to the cross-section to Lyman-band emitting states \cite{Janev}. Moreover, the difference in Lyman-alpha emission between present results and low-temperature sources \cite{Komppula_exp} probably correlates to the plasma opacity. It depends on the hydrogen atom density, which is high according to molecular continuum emission.

The obtained results warrant considering the quasi-dynamic ECR discharge VUV source as a unique facility combining advantages of various experimental approaches. One can obtain much higher volumetric VUV power, than with the use of other experimental RF discharges, or reach the same order in efficiency in comparison with conventional VUV sources (e.g., excimer lamps), but with much higher total VUV power \cite{Park}. 

\begin{acknowledgments}
The work was carried out with the support of RFBR grant $\#$20-32-70002 and within the framework of state assignment no. 0035-2019-0002.
\end{acknowledgments}

\section*{Data Availability}
The data that support the findings of this study are available from the corresponding author upon reasonable request.

\bibliography{Lapin}

\providecommand{\noopsort}[1]{}\providecommand{\singleletter}[1]{#1}%
\begin{thebibliography}{21}%
\makeatletter
\providecommand \@ifxundefined [1]{%
 \@ifx{#1\undefined}
}%
\providecommand \@ifnum [1]{%
 \ifnum #1\expandafter \@firstoftwo
 \else \expandafter \@secondoftwo
 \fi
}%
\providecommand \@ifx [1]{%
 \ifx #1\expandafter \@firstoftwo
 \else \expandafter \@secondoftwo
 \fi
}%
\providecommand \natexlab [1]{#1}%
\providecommand \enquote  [1]{``#1''}%
\providecommand \bibnamefont  [1]{#1}%
\providecommand \bibfnamefont [1]{#1}%
\providecommand \citenamefont [1]{#1}%
\providecommand \href@noop [0]{\@secondoftwo}%
\providecommand \href [0]{\begingroup \@sanitize@url \@href}%
\providecommand \@href[1]{\@@startlink{#1}\@@href}%
\providecommand \@@href[1]{\endgroup#1\@@endlink}%
\providecommand \@sanitize@url [0]{\catcode `\\12\catcode `\$12\catcode
  `\&12\catcode `\#12\catcode `\^12\catcode `\_12\catcode `\%12\relax}%
\providecommand \@@startlink[1]{}%
\providecommand \@@endlink[0]{}%
\providecommand \url  [0]{\begingroup\@sanitize@url \@url }%
\providecommand \@url [1]{\endgroup\@href {#1}{\urlprefix }}%
\providecommand \urlprefix  [0]{URL }%
\providecommand \Eprint [0]{\href }%
\providecommand \doibase [0]{http://dx.doi.org/}%
\providecommand \selectlanguage [0]{\@gobble}%
\providecommand \bibinfo  [0]{\@secondoftwo}%
\providecommand \bibfield  [0]{\@secondoftwo}%
\providecommand \translation [1]{[#1]}%
\providecommand \BibitemOpen [0]{}%
\providecommand \bibitemStop [0]{}%
\providecommand \bibitemNoStop [0]{.\EOS\space}%
\providecommand \EOS [0]{\spacefactor3000\relax}%
\providecommand \BibitemShut  [1]{\csname bibitem#1\endcsname}%
\let\auto@bib@innerbib\@empty
\bibitem [{\citenamefont {Schneider}\ \emph {et~al.}(2005)\citenamefont
  {Schneider}, \citenamefont {Sanchez},\ and\ \citenamefont
  {Achard}}]{Schneider}%
  \BibitemOpen
  \bibfield  {author} {\bibinfo {author} {\bibfnamefont {H.}~\bibnamefont
  {Schneider}}, \bibinfo {author} {\bibfnamefont {J.-L.}\ \bibnamefont
  {Sanchez}}, \ and\ \bibinfo {author} {\bibfnamefont {J.}~\bibnamefont
  {Achard}},\ }in\ \href@noop {} {\emph {\bibinfo {booktitle} {2005 European
  Conference on Power Electronics and Applications}}}\ (\bibinfo  {publisher}
  {IEEE Service Center},\ \bibinfo {address} {Dresden, Germany},\ \bibinfo
  {year} {2005})\ p.~\bibinfo {pages} {9}\BibitemShut {NoStop}%
\bibitem [{\citenamefont {Calvani}\ \emph {et~al.}(2011)\citenamefont
  {Calvani}, \citenamefont {Girolami}, \citenamefont {Ricciotti},\ and\
  \citenamefont {Conte}}]{Calvani}%
  \BibitemOpen
  \bibfield  {author} {\bibinfo {author} {\bibfnamefont {P.}~\bibnamefont
  {Calvani}}, \bibinfo {author} {\bibfnamefont {M.}~\bibnamefont {Girolami}},
  \bibinfo {author} {\bibfnamefont {G.}~\bibnamefont {Ricciotti}}, \ and\
  \bibinfo {author} {\bibfnamefont {G.}~\bibnamefont {Conte}},\ }in\ \href@noop
  {} {\emph {\bibinfo {booktitle} {Proc. SPIE 8069, Integrated Photonics:
  Materials, Devices, and Applications}}}\ (\bibinfo  {publisher} {SPIE},\
  \bibinfo {address} {Prague, Czech Republic},\ \bibinfo {year} {2011})\ p.\
  \bibinfo {pages} {806908}\BibitemShut {NoStop}%
\bibitem [{\citenamefont {Girolami}\ \emph {et~al.}(2011)\citenamefont
  {Girolami}, \citenamefont {Galbiati},\ and\ \citenamefont
  {Salvatori}}]{Girolami}%
  \BibitemOpen
  \bibfield  {author} {\bibinfo {author} {\bibfnamefont {M.}~\bibnamefont
  {Girolami}}, \bibinfo {author} {\bibfnamefont {A.}~\bibnamefont {Galbiati}},
  \ and\ \bibinfo {author} {\bibfnamefont {S.}~\bibnamefont {Salvatori}},\ }in\
  \href@noop {} {\emph {\bibinfo {booktitle} {2011 IEEE Nuclear Science
  Symposium Conference Record}}}\ (\bibinfo  {publisher} {IEEE},\ \bibinfo
  {address} {Valencia, Spain},\ \bibinfo {year} {2011})\ pp.\ \bibinfo {pages}
  {586--590}\BibitemShut {NoStop}%
\bibitem [{\citenamefont {Geis}\ \emph {et~al.}(2004)\citenamefont {Geis},
  \citenamefont {Krohn}, \citenamefont {Lawless}, \citenamefont {Deneault},
  \citenamefont {Marchant}, \citenamefont {Twichell}, \citenamefont
  {Lyszczarz}, \citenamefont {Butler}, \citenamefont {Flechtner},\ and\
  \citenamefont {Wright}}]{Geis}%
  \BibitemOpen
  \bibfield  {author} {\bibinfo {author} {\bibfnamefont {M.~W.}\ \bibnamefont
  {Geis}}, \bibinfo {author} {\bibfnamefont {K.~E.}\ \bibnamefont {Krohn}},
  \bibinfo {author} {\bibfnamefont {J.~M.}\ \bibnamefont {Lawless}}, \bibinfo
  {author} {\bibfnamefont {S.~J.}\ \bibnamefont {Deneault}}, \bibinfo {author}
  {\bibfnamefont {M.~F.}\ \bibnamefont {Marchant}}, \bibinfo {author}
  {\bibfnamefont {J.~C.}\ \bibnamefont {Twichell}}, \bibinfo {author}
  {\bibfnamefont {T.~M.}\ \bibnamefont {Lyszczarz}}, \bibinfo {author}
  {\bibfnamefont {J.~E.}\ \bibnamefont {Butler}}, \bibinfo {author}
  {\bibfnamefont {D.~D.}\ \bibnamefont {Flechtner}}, \ and\ \bibinfo {author}
  {\bibfnamefont {R.}~\bibnamefont {Wright}},\ }\href@noop {} {\bibfield
  {journal} {\bibinfo  {journal} {Appl. Phys. Lett.}\ }\textbf {\bibinfo
  {volume} {84}},\ \bibinfo {pages} {4620} (\bibinfo {year}
  {2004})}\BibitemShut {NoStop}%
\bibitem [{\citenamefont {Yatsenko}\ \emph {et~al.}(2005)\citenamefont
  {Yatsenko}, \citenamefont {Bachau}, \citenamefont {Belsky}, \citenamefont
  {Gaudin}, \citenamefont {Geoffroy}, \citenamefont {Guizard}, \citenamefont
  {Martin}, \citenamefont {Petite}, \citenamefont {Philippov},\ and\
  \citenamefont {Vasil’ev}}]{Yatsenko}%
  \BibitemOpen
  \bibfield  {author} {\bibinfo {author} {\bibfnamefont {B.~N.}\ \bibnamefont
  {Yatsenko}}, \bibinfo {author} {\bibfnamefont {H.}~\bibnamefont {Bachau}},
  \bibinfo {author} {\bibfnamefont {A.~N.}\ \bibnamefont {Belsky}}, \bibinfo
  {author} {\bibfnamefont {J.}~\bibnamefont {Gaudin}}, \bibinfo {author}
  {\bibfnamefont {G.}~\bibnamefont {Geoffroy}}, \bibinfo {author}
  {\bibfnamefont {S.}~\bibnamefont {Guizard}}, \bibinfo {author} {\bibfnamefont
  {P.}~\bibnamefont {Martin}}, \bibinfo {author} {\bibfnamefont
  {G.}~\bibnamefont {Petite}}, \bibinfo {author} {\bibfnamefont
  {A.}~\bibnamefont {Philippov}}, \ and\ \bibinfo {author} {\bibfnamefont
  {A.~N.}\ \bibnamefont {Vasil’ev}},\ }\href@noop {} {\bibfield  {journal}
  {\bibinfo  {journal} {Phys. Stat. Sol.}\ }\textbf {\bibinfo {volume} {2}},\
  \bibinfo {pages} {240} (\bibinfo {year} {2005})}\BibitemShut {NoStop}%
\bibitem [{\citenamefont {Park}\ \emph {et~al.}(2017)\citenamefont {Park},
  \citenamefont {Herring}, \citenamefont {Mironov}, \citenamefont {Cho},\ and\
  \citenamefont {Eden}}]{Park}%
  \BibitemOpen
  \bibfield  {author} {\bibinfo {author} {\bibfnamefont {S.-J.}\ \bibnamefont
  {Park}}, \bibinfo {author} {\bibfnamefont {C.~M.}\ \bibnamefont {Herring}},
  \bibinfo {author} {\bibfnamefont {A.~E.}\ \bibnamefont {Mironov}}, \bibinfo
  {author} {\bibfnamefont {J.~H.}\ \bibnamefont {Cho}}, \ and\ \bibinfo
  {author} {\bibfnamefont {J.~G.}\ \bibnamefont {Eden}},\ }\href@noop {}
  {\bibfield  {journal} {\bibinfo  {journal} {APL Photonics}\ }\textbf
  {\bibinfo {volume} {2}},\ \bibinfo {pages} {041302} (\bibinfo {year}
  {2017})}\BibitemShut {NoStop}%
\bibitem [{\citenamefont {Ajello}\ \emph {et~al.}(1982)\citenamefont {Ajello},
  \citenamefont {Srivastava},\ and\ \citenamefont {Yung}}]{Ajello}%
  \BibitemOpen
  \bibfield  {author} {\bibinfo {author} {\bibfnamefont {J.~M.}\ \bibnamefont
  {Ajello}}, \bibinfo {author} {\bibfnamefont {S.~K.}\ \bibnamefont
  {Srivastava}}, \ and\ \bibinfo {author} {\bibfnamefont {Y.~L.}\ \bibnamefont
  {Yung}},\ }\href@noop {} {\bibfield  {journal} {\bibinfo  {journal} {Phys.
  Rev. A}\ }\textbf {\bibinfo {volume} {25}},\ \bibinfo {pages} {2485}
  (\bibinfo {year} {1982})}\BibitemShut {NoStop}%
\bibitem [{\citenamefont {Komppula}\ and\ \citenamefont
  {Tarvainen}(2015{\natexlab{a}})}]{Komppula_th}%
  \BibitemOpen
  \bibfield  {author} {\bibinfo {author} {\bibfnamefont {J.}~\bibnamefont
  {Komppula}}\ and\ \bibinfo {author} {\bibfnamefont {O.}~\bibnamefont
  {Tarvainen}},\ }\href@noop {} {\bibfield  {journal} {\bibinfo  {journal}
  {Phys. Plasmas}\ }\textbf {\bibinfo {volume} {22}},\ \bibinfo {pages}
  {103516} (\bibinfo {year} {2015}{\natexlab{a}})}\BibitemShut {NoStop}%
\bibitem [{\citenamefont {Batishchev}\ and\ \citenamefont
  {Molvig}(2000)}]{Batishchev}%
  \BibitemOpen
  \bibfield  {author} {\bibinfo {author} {\bibfnamefont {O.}~\bibnamefont
  {Batishchev}}\ and\ \bibinfo {author} {\bibfnamefont {K.}~\bibnamefont
  {Molvig}},\ }in\ \href@noop {} {\emph {\bibinfo {booktitle} {36th
  AIAA/ASME/SAE/ASEE Joint Propulsion Conference and Exhibit}}}\ (\bibinfo
  {publisher} {IEEE},\ \bibinfo {address} {Las Vegas, USA},\ \bibinfo {year}
  {2000})\ p.\ \bibinfo {pages} {3754}\BibitemShut {NoStop}%
\bibitem [{\citenamefont {Komppula}\ \emph {et~al.}(2015)\citenamefont
  {Komppula}, \citenamefont {Tarvainen}, \citenamefont {Kalvas}, \citenamefont
  {Koivisto}, \citenamefont {Kronholm}, \citenamefont {Laulainen},\ and\
  \citenamefont {Myllyperki{\"o}}}]{Komppula_exp}%
  \BibitemOpen
  \bibfield  {author} {\bibinfo {author} {\bibfnamefont {J.}~\bibnamefont
  {Komppula}}, \bibinfo {author} {\bibfnamefont {O.}~\bibnamefont {Tarvainen}},
  \bibinfo {author} {\bibfnamefont {T.}~\bibnamefont {Kalvas}}, \bibinfo
  {author} {\bibfnamefont {H.}~\bibnamefont {Koivisto}}, \bibinfo {author}
  {\bibfnamefont {R.}~\bibnamefont {Kronholm}}, \bibinfo {author}
  {\bibfnamefont {J.}~\bibnamefont {Laulainen}}, \ and\ \bibinfo {author}
  {\bibfnamefont {P.}~\bibnamefont {Myllyperki{\"o}}},\ }\href@noop {}
  {\bibfield  {journal} {\bibinfo  {journal} {J. Phys. D: Appl. Phys.}\
  }\textbf {\bibinfo {volume} {48}},\ \bibinfo {pages} {365201} (\bibinfo
  {year} {2015})}\BibitemShut {NoStop}%
\bibitem [{\citenamefont {Geller}(1998)}]{Geller}%
  \BibitemOpen
  \bibfield  {author} {\bibinfo {author} {\bibfnamefont {R.}~\bibnamefont
  {Geller}},\ }\href@noop {} {\bibfield  {journal} {\bibinfo  {journal} {Rev.
  Sci. Instrum.}\ }\textbf {\bibinfo {volume} {69}},\ \bibinfo {pages} {1302}
  (\bibinfo {year} {1998})}\BibitemShut {NoStop}%
\bibitem [{\citenamefont {Sun}\ \emph {et~al.}(2020)\citenamefont {Sun},
  \citenamefont {Zhao}, \citenamefont {Zha}, \citenamefont {Lu}, \citenamefont
  {Guo}, \citenamefont {Cao}, \citenamefont {Wu}, \citenamefont {Qian},
  \citenamefont {Yang}, \citenamefont {Fang}, \citenamefont {Zhang},
  \citenamefont {Zhang}, \citenamefont {Guo},\ and\ \citenamefont {Liu}}]{Sun}%
  \BibitemOpen
  \bibfield  {author} {\bibinfo {author} {\bibfnamefont {L.}~\bibnamefont
  {Sun}}, \bibinfo {author} {\bibfnamefont {H.~W.}\ \bibnamefont {Zhao}},
  \bibinfo {author} {\bibfnamefont {H.~Y.}\ \bibnamefont {Zha}}, \bibinfo
  {author} {\bibfnamefont {W.}~\bibnamefont {Lu}}, \bibinfo {author}
  {\bibfnamefont {J.~W.}\ \bibnamefont {Guo}}, \bibinfo {author} {\bibfnamefont
  {Y.}~\bibnamefont {Cao}}, \bibinfo {author} {\bibfnamefont {Q.}~\bibnamefont
  {Wu}}, \bibinfo {author} {\bibfnamefont {C.}~\bibnamefont {Qian}}, \bibinfo
  {author} {\bibfnamefont {Y.}~\bibnamefont {Yang}}, \bibinfo {author}
  {\bibfnamefont {X.}~\bibnamefont {Fang}}, \bibinfo {author} {\bibfnamefont
  {Z.~M.}\ \bibnamefont {Zhang}}, \bibinfo {author} {\bibfnamefont {X.~Z.}\
  \bibnamefont {Zhang}}, \bibinfo {author} {\bibfnamefont {X.~H.}\ \bibnamefont
  {Guo}}, \ and\ \bibinfo {author} {\bibfnamefont {Z.~W.}\ \bibnamefont
  {Liu}},\ }\href@noop {} {\bibfield  {journal} {\bibinfo  {journal} {Rev. Sci.
  Instrum.}\ }\textbf {\bibinfo {volume} {91}},\ \bibinfo {pages} {023310}
  (\bibinfo {year} {2020})}\BibitemShut {NoStop}%
\bibitem [{\citenamefont {Golubev}\ \emph {et~al.}(2004)\citenamefont
  {Golubev}, \citenamefont {Razin}, \citenamefont {Sidorov}, \citenamefont
  {Skalyga}, \citenamefont {Vodopyanov},\ and\ \citenamefont
  {Zorin}}]{Golubev04}%
  \BibitemOpen
  \bibfield  {author} {\bibinfo {author} {\bibfnamefont {S.~V.}\ \bibnamefont
  {Golubev}}, \bibinfo {author} {\bibfnamefont {S.~V.}\ \bibnamefont {Razin}},
  \bibinfo {author} {\bibfnamefont {A.~V.}\ \bibnamefont {Sidorov}}, \bibinfo
  {author} {\bibfnamefont {V.~A.}\ \bibnamefont {Skalyga}}, \bibinfo {author}
  {\bibfnamefont {A.~V.}\ \bibnamefont {Vodopyanov}}, \ and\ \bibinfo {author}
  {\bibfnamefont {V.}~\bibnamefont {Zorin}},\ }\href@noop {} {\bibfield
  {journal} {\bibinfo  {journal} {Rev. Sci. Instrum.}\ }\textbf {\bibinfo
  {volume} {75}},\ \bibinfo {pages} {1675} (\bibinfo {year}
  {2004})}\BibitemShut {NoStop}%
\bibitem [{\citenamefont {Skalyga}\ \emph {et~al.}(2014)\citenamefont
  {Skalyga}, \citenamefont {Izotov}, \citenamefont {Razin}, \citenamefont
  {Sidorov}, \citenamefont {Golubev}, \citenamefont {Kalvas}, \citenamefont
  {Koivisto},\ and\ \citenamefont {Tarvainen}}]{Skalyga2014}%
  \BibitemOpen
  \bibfield  {author} {\bibinfo {author} {\bibfnamefont {V.}~\bibnamefont
  {Skalyga}}, \bibinfo {author} {\bibfnamefont {I.}~\bibnamefont {Izotov}},
  \bibinfo {author} {\bibfnamefont {S.}~\bibnamefont {Razin}}, \bibinfo
  {author} {\bibfnamefont {A.}~\bibnamefont {Sidorov}}, \bibinfo {author}
  {\bibfnamefont {S.}~\bibnamefont {Golubev}}, \bibinfo {author} {\bibfnamefont
  {T.}~\bibnamefont {Kalvas}}, \bibinfo {author} {\bibfnamefont
  {H.}~\bibnamefont {Koivisto}}, \ and\ \bibinfo {author} {\bibfnamefont
  {O.}~\bibnamefont {Tarvainen}},\ }\href@noop {} {\bibfield  {journal}
  {\bibinfo  {journal} {Rev. Sci. Instrum.}\ }\textbf {\bibinfo {volume}
  {85}},\ \bibinfo {pages} {02A702} (\bibinfo {year} {2014})}\BibitemShut
  {NoStop}%
\bibitem [{\citenamefont {Skalyga}\ \emph {et~al.}(2016)\citenamefont
  {Skalyga}, \citenamefont {Izotov}, \citenamefont {Razin}, \citenamefont
  {Sidorov}, \citenamefont {Golubev}, \citenamefont {Kalvas}, \citenamefont
  {Koivisto},\ and\ \citenamefont {Tarvainen}}]{Skalyga2016}%
  \BibitemOpen
  \bibfield  {author} {\bibinfo {author} {\bibfnamefont {V.}~\bibnamefont
  {Skalyga}}, \bibinfo {author} {\bibfnamefont {I.}~\bibnamefont {Izotov}},
  \bibinfo {author} {\bibfnamefont {S.}~\bibnamefont {Razin}}, \bibinfo
  {author} {\bibfnamefont {A.}~\bibnamefont {Sidorov}}, \bibinfo {author}
  {\bibfnamefont {S.}~\bibnamefont {Golubev}}, \bibinfo {author} {\bibfnamefont
  {T.}~\bibnamefont {Kalvas}}, \bibinfo {author} {\bibfnamefont
  {H.}~\bibnamefont {Koivisto}}, \ and\ \bibinfo {author} {\bibfnamefont
  {O.}~\bibnamefont {Tarvainen}},\ }\href@noop {} {\bibfield  {journal}
  {\bibinfo  {journal} {Rev. Sci. Instrum.}\ }\textbf {\bibinfo {volume}
  {87}},\ \bibinfo {pages} {02A716} (\bibinfo {year} {2016})}\BibitemShut
  {NoStop}%
\bibitem [{\citenamefont {Skalyga}\ \emph {et~al.}(2017)\citenamefont
  {Skalyga}, \citenamefont {Izotov}, \citenamefont {Sidorov}, \citenamefont
  {Golubev},\ and\ \citenamefont {Razin}}]{Skalyga2017}%
  \BibitemOpen
  \bibfield  {author} {\bibinfo {author} {\bibfnamefont {V.~A.}\ \bibnamefont
  {Skalyga}}, \bibinfo {author} {\bibfnamefont {I.~V.}\ \bibnamefont {Izotov}},
  \bibinfo {author} {\bibfnamefont {A.~V.}\ \bibnamefont {Sidorov}}, \bibinfo
  {author} {\bibfnamefont {S.~V.}\ \bibnamefont {Golubev}}, \ and\ \bibinfo
  {author} {\bibfnamefont {S.~V.}\ \bibnamefont {Razin}},\ }\href@noop {}
  {\bibfield  {journal} {\bibinfo  {journal} {Rev. Sci. Instrum.}\ }\textbf
  {\bibinfo {volume} {88}},\ \bibinfo {pages} {033503} (\bibinfo {year}
  {2017})}\BibitemShut {NoStop}%
\bibitem [{\citenamefont {Janev}\ \emph {et~al.}(2003)\citenamefont {Janev},
  \citenamefont {Reiter},\ and\ \citenamefont {Samm}}]{Janev}%
  \BibitemOpen
  \bibfield  {author} {\bibinfo {author} {\bibfnamefont {R.~K.}\ \bibnamefont
  {Janev}}, \bibinfo {author} {\bibfnamefont {D.}~\bibnamefont {Reiter}}, \
  and\ \bibinfo {author} {\bibfnamefont {U.}~\bibnamefont {Samm}},\ }\href@noop
  {} {\emph {\bibinfo {title} {Collision processes in low-temperature hydrogen
  plasmas}}},\ \bibinfo {series} {Berichte des Forschungszentrums Jülich},
  Vol.\ \bibinfo {volume} {JUEL-4105}\ (\bibinfo  {publisher}
  {Forschungszentrum, Zentralbibliothek},\ \bibinfo {year} {2003})\BibitemShut
  {NoStop}%
\bibitem [{\citenamefont {Dijkstra}(2017)}]{Dijkstra}%
  \BibitemOpen
  \bibfield  {author} {\bibinfo {author} {\bibfnamefont {M.}~\bibnamefont
  {Dijkstra}},\ }\href@noop {} {\enquote {\bibinfo {title} {Physics of
  {Ly}$_\alpha$ radiative transfer},}\ }\bibinfo {howpublished} {e-print arXiv:
  astro-ph.GA/1704.03416v1} (\bibinfo {year} {2017})\BibitemShut {NoStop}%
\bibitem [{\citenamefont {Komppula}\ and\ \citenamefont
  {Tarvainen}(2015{\natexlab{b}})}]{Komppula_exp_1}%
  \BibitemOpen
  \bibfield  {author} {\bibinfo {author} {\bibfnamefont {J.}~\bibnamefont
  {Komppula}}\ and\ \bibinfo {author} {\bibfnamefont {O.}~\bibnamefont
  {Tarvainen}},\ }\href@noop {} {\bibfield  {journal} {\bibinfo  {journal}
  {Plasma Sources Sci. Technol}\ }\textbf {\bibinfo {volume} {24}},\ \bibinfo
  {pages} {045008} (\bibinfo {year} {2015}{\natexlab{b}})}\BibitemShut
  {NoStop}%
\bibitem [{\citenamefont {Skalyga}\ \emph {et~al.}(2012)\citenamefont
  {Skalyga}, \citenamefont {Izotov}, \citenamefont {Sidorov}, \citenamefont
  {Razin}, \citenamefont {Zorin}, \citenamefont {Tarvainen}, \citenamefont
  {Koivisto},\ and\ \citenamefont {Kalvas}}]{Skalyga2012}%
  \BibitemOpen
  \bibfield  {author} {\bibinfo {author} {\bibfnamefont {V.}~\bibnamefont
  {Skalyga}}, \bibinfo {author} {\bibfnamefont {I.}~\bibnamefont {Izotov}},
  \bibinfo {author} {\bibfnamefont {A.}~\bibnamefont {Sidorov}}, \bibinfo
  {author} {\bibfnamefont {S.}~\bibnamefont {Razin}}, \bibinfo {author}
  {\bibfnamefont {V.}~\bibnamefont {Zorin}}, \bibinfo {author} {\bibfnamefont
  {O.}~\bibnamefont {Tarvainen}}, \bibinfo {author} {\bibfnamefont
  {H.}~\bibnamefont {Koivisto}}, \ and\ \bibinfo {author} {\bibfnamefont
  {T.}~\bibnamefont {Kalvas}},\ }\href@noop {} {\bibfield  {journal} {\bibinfo
  {journal} {JINST}\ }\textbf {\bibinfo {volume} {5}},\ \bibinfo {pages}
  {P10010} (\bibinfo {year} {2012})}\BibitemShut {NoStop}%
\bibitem [{\citenamefont {Behringer}\ and\ \citenamefont
  {Fantz}(2000)}]{Behringer}%
  \BibitemOpen
  \bibfield  {author} {\bibinfo {author} {\bibfnamefont {K.}~\bibnamefont
  {Behringer}}\ and\ \bibinfo {author} {\bibfnamefont {U.}~\bibnamefont
  {Fantz}},\ }\href@noop {} {\bibfield  {journal} {\bibinfo  {journal} {New J.
  Phys.}\ }\textbf {\bibinfo {volume} {2}},\ \bibinfo {pages} {23} (\bibinfo
  {year} {2000})}\BibitemShut {NoStop}%
\end{thebibliography}%


\providecommand{\noopsort}[1]{}\providecommand{\singleletter}[1]{#1}%
%

\end{document}